\magnification1020
\input epsf.sty
\input g.sty

\hsize13.97truecm
\vsize19.1truecm
\hyphenation{Akh-met-shin}

\nopagenumbers
\rightline\timestamp
\rightline{FTUAM 04-02}
%\rightline{February, 2004}
\rightline{hep-ph/0402285}
\bigskip
\hrule height .3mm
\vskip.6cm
\centerline{{\bigfib The hadronic contributions to the  anomalous}}
\medskip\centerline{{\bigfib magnetic moment of the muon}}
\medskip
\centerrule{.7cm}
\vskip1cm

\setbox9=\vbox{\hsize65mm {\noindent\fib J. F. de Troc\'oniz and  F. J. 
Yndur\'ain} 
\vskip .1cm
\noindent{\addressfont Departamento de F\'{\i}sica Te\'orica, C-XI\hb
 Universidad Aut\'onoma de Madrid,\hb
 Canto Blanco,\hb
E-28049, Madrid, Spain.}\hb}
\smallskip
\centerline{\box9}
\bigskip
\setbox0=\vbox{\abstracttype{Abstract}  We present a new, completely revised 
calculation of the muon anomalous magnetic moment, $a_\mu=(g_{\mu}-2)/2$, 
 comparing it with the more recent experimental determination of this quantity; this furnishes 
 an important test of theories of strong, weak and electromagnetic interactions. 
These theoretical and experimental determinations give the very precise numbers, 
$$10^{11}\times a_\mu=\cases{
116\,591\,806\pm50\pm10\;({\rm rad.})\pm30\;(\ell\times\ell)\quad\hbox{[Th., no $\tau$]}\cr
116\,591\,889\pm49\pm10\;({\rm rad.})\pm30\;(\ell\times\ell)\quad\hbox{[Theory, $\tau$]}\cr
116\,592\,080\pm60\quad\hbox{[Experiment]}.\cr}$$ 
In the theoretical evaluations, the first quantity does not, and the second one does, 
use information from $\tau$ decay. The first errors for the theoretical evaluations 
include statistical plus systematic errors; the  other ones are the estimated errors due to incomplete
treatment of  radiative corrections and the estimated error in the light-by-light
 scattering contribution. We thus have a significant mismatch 
between theory and experiment. We also use part of the theoretical calculations
to give a precise  evaluation of  the electromagnetic coupling on the $Z$, $\bar{\alpha}_{\rm Q.E.D.}(M^2_{Z})$,
of the masses and widths of
 the (charged and neutral) rho resonances, of the scattering length and effective range  for the P wave in
$\pi\pi$ scattering, and of the quadratic radius and second coefficient of the pion form factor. }
\centerline{\box0}
\vfill\eject

\brochureb{\smallsc the hadronic contributions
 to the anomalous magnetic moment
of the muon}{\smallsc  j. f. de troc\'oniz and f. j.  yndur\'ain}{1}
\brochureendcover{Typeset with \physmatex}

\booksection{1. Introduction}

\noindent
The anomalous magnetic moments of electrons and muons 
provide one of the more impressive tests of the 
standard model of strong, weak and electromagnetic interactions. 
The electron anomaly, $a_e$, receives only marginal contributions 
from weak and strong interactions, 
being dominated by electromagnetic (QED) radiative 
corrections. 
The agreement between theory and experiment is such that, if we turn it around, 
the experimental value of $a_e$ 
provides the more precise determination of the fine structure constant,\ref{1} $\alpha$.

For the muon magnetic moment, the recent precise measurements\ref{2} 
of the muon anomaly, $a_\mu$,
constitute one of the more impressive tests not only of electroweak interactions, 
but of strong interactions as well. 
After these measurements, the world average value for $a_\mu=(g_{\mu}-2)/2$ is 
$$10^{11}\times a_\mu=116\,592\,080\pm60\quad\hbox{[Experiment]}.
\eqno{(1.1)}
$$
Considering that theory also gives the value of the magnetic moment itself 
(and not only the anomaly) the agreement of theory and experiment 
that we will describe represents a precision of one or two parts in a {\sl billion}.
Nevertheless, and as we will see, there remains a discrepancy at the level of 
$2.3\,\sigma$ to $3.3\,\sigma$.

The electromagnetic and weak contributions to $a_\mu$ have been 
calculated with great accuracy:\ref{3}
$$\eqalign{
10^{11}\times a_\mu(\hbox{QED})=&\,116\,584\,719\pm1.8,\cr
10^{11}\times a_\mu(\hbox{Weak})=&\,\phantom{116\,584\,}152\pm3.
\cr}
\eqno{(1.2)}$$
Combining this with (1.1), we find the {\sl experimental} 
number for the hadronic contributions to $a_\mu$,
$$
10^{11}\times a_\mu(\hbox{Hadr.})=7\,209\pm60\quad\hbox{[Experiment]}.
\eqno{(1.3)}$$
The evaluation of this quantity, $a_\mu(\hbox{Hadr.})$, from 
theory will be the main subject of the present note; we will find 
$$10^{11}\times a_\mu(\hbox{Hadr.})=\cases{
6\,935\pm50\pm10\;({\rm rad.})\pm30\,(\ell\times\ell)\quad\hbox{[No $\tau$]}\cr
7\,018\pm49\pm10\,({\rm rad.})\pm30\,(\ell\times\ell)
\quad\hbox{[With $\tau$]}.\cr
}
\equn{(1.4)}$$
The numbers above depend on whether or not one includes information on 
$\tau$ decay, which is probably  
the more reliable result as it is the one that incorporates more information: 
see our text
below for details. In the errors,  ``rad" and ``$\ell\times\ell$" refer, respectively, to
estimated errors due to  uncalculated radiative corrections and light-by-light scattering
contributions. The results in (1.4) takes into account all the more recent  
$e^+e^-$ annihilations data, and $e\pi$ scattering data. 
At the end of the present article we 
will comment on the degree of   agreement of theory and 
experiment, compare our results with those of other 
recent calculations, and  discuss the 
possible reasons for the  discrepancy between (1.3) and (1.4).
 
We will also give a summary of the results that some of the calculations imply 
 for   
the electromagnetic coupling on the $Z$, $\bar{\alpha}_{\rm Q.E.D.}(M^2_{Z})$, 
for the masses and widths of the (charged and neutral) rho 
resonances, for the scattering length and effective radius for the P wave in pion-pion scattering,
 and for the quadratic
radius and second coefficient of the pion form factor.

\booksection{2. The hadronic contributions to $a_\mu$. I: the $O(\alpha^2)$ piece}

\noindent
To order $\alpha^2$, the contributions to $a_\mu(\hbox{Hadr.})$ 
can be represented by the diagram shown in 
\fig~1. As has been 
known for a long time, they can be written in terms of the 
cross section for $e^+e^-$ annihilation into hadrons, as follows. 
We first write 
$$\eqalign{
a^{(2)}_\mu(\hbox{Hadr.})=&12\pi\int_{4m^2_\pi}^\infty \dd s\,K(s)\imag\piv(s),\cr
K(s)=&\dfrac{\alpha^2}{3\pi^2s}\hat{K}(s);\quad \hat{K}(s)=
\int_0^1\dd x\,\dfrac{x^2(1-x)}{x^2+(1-x)s/m^2_\mu}.\cr}
\equn{(2.1a)}$$
Here $\piv$ is the hadronic part of the photon 
vacuum polarization function. 
Then, we can express
 $\imag \Piv$ in terms of 
the ratio of (lowest order) cross sections for $e^+e^-$ annihilation into hadrons 
over annihilation into muons:
$$R(s)=\dfrac{\sigma^{(0)}(e^+e^-\to{\rm hadrons})}{\sigma^{(0)}(e^+e^-\to\mu^+\mu^-)},
\quad \sigma^{(0)}(e^+e^-\to\mu^+\mu^-)\equiv\dfrac{4\pi\alpha^2}{3s}:$$
$$a^{(2)}_\mu(\hbox{Hadr.})=\int_{4m^2_\pi}^\infty \dd s\,K(s) R(s).
\equn{(2.1b)}$$
 
\topinsert{
\setbox0=\vbox{\hsize8.4truecm{\epsfxsize 7.truecm\epsfbox{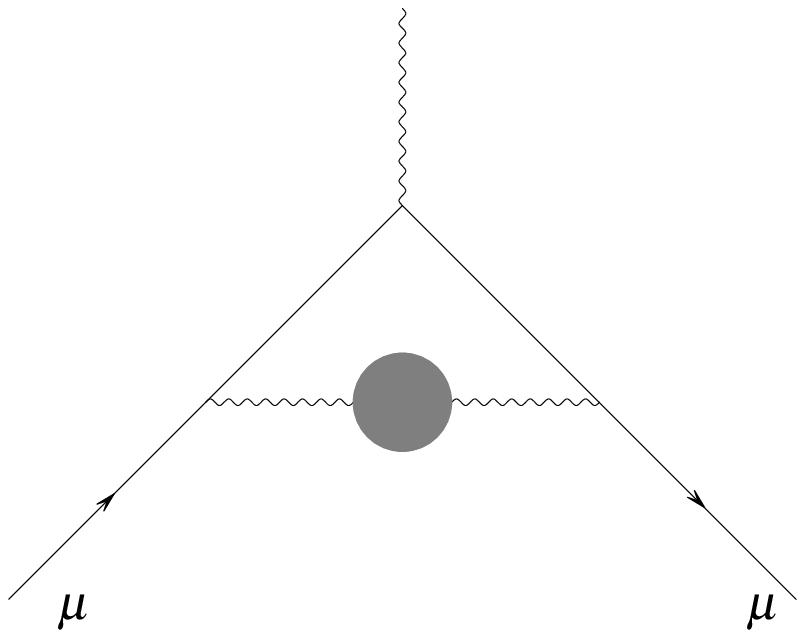}}} 
\setbox6=\vbox{\hsize 4.5truecm\captiontype\figurasc{Figure 1. }{The order $\alpha^2$
hadronic contributions to the muon magnetic moment. 
The blob represents an arbitrary hadronic state. 
The wavy lines are photons.\hb
\phantom{XX}}\hb
\vskip.1cm} 
\medskip
\line{\box0\hfil\box6}
\medskip
}\endinsert

Therefore, the situation is, in principle, simple: we take the {\sl experimental} 
cross section for $e^+e^-\to{\rm hadrons}$, insert it into (2.1b) and the value for 
$a^{(2)}_\mu(\hbox{Hadr.})$ will follow.
In practice, however,  things are more complex. 
The experimental  numbers for the cross section $e^+e^-$ into hadrons   
present (relatively) large errors and  we have, therefore,
 interest to supplement these with more precise theoretical
formulas  whenever possible. 
As a matter of fact, by so doing we are able to 
diminish the error in the theoretical calculation almost by a factor of two. 

\vfill
\booksubsection{2.1 The low energy region, $s\leq0.8\,\gev^2$}

\noindent
In the region below $s=0.8\,\gev^2$, that we may call ``rho region", experimental data 
 have in fact improved substantially in the last years due to the more recent measurements, 
especially at Novosibirsk.\ref{4,5}  
Here the important contributions are those of the omega resonance 
(for which experimental data and the Gounnaris--Sakurai method may 
be used; see  below  and ref.~6) and the two-pion contribution, 
which is the one that may be made more precise using theory, and 
also the one that we will discuss 
in more detail because it provides the bulk 
of $a_\mu({\rm Hadr.})$. 
Here one profits from the fact that the 
 two pion contribution  can be expressed in terms of the 
pion form factor, $F_\pi$,
$$\imag \piv_{2\pi}(s)=\dfrac{1}{48\pi}\left(1-\dfrac{4m^2_\pi}{s}\right)^{3/2} 
|F_\pi(s)|^2,\equn{(2.2)}$$
and $F_\pi$ may be determined from fits to data on $e^+e^-\to2\pi$ and, 
using analyticity, also data from $\pi e\to\pi e$ scattering, i.e., at spacelike $s$. 
What is more, we may use (with due caution; see below)
 data on $\tau$ decay, $\tau^\pm\to \nu \pi^\pm\pi^0$, 
related to $F_\pi$ by isospin invariance.

We will for the moment work in the approximation of neglecting electroweak corrections to $F_\pi$; 
we will discuss this further in \subsect~2.2. 
In this approximation, 
the properties of $F_\pi(s)$ that allow us an improved calculation are the following: 
\item{(i) }{$F_\pi(s)$ is an analytic function of $s$, with a cut 
from $4m^2_\pi$ to infinity.}
\item{(ii) }{On the cut, the phase of $F_\pi(s)$ is, because of unitarity, identical to 
that of the P-wave in $\pi\pi$ scattering, $\delta_1^1(s)$, and 
this equality 
holds until the opening of the inelastic threshold at $s=t_0$
 (Fermi--Watson final state interaction 
theorem).}
\item{(iii) }{For large $s$, $F_\pi(s)\simeq 1/s$. Actually, one knows the coefficient of this 
behaviour, but we will not need it here.}
\item{(iv) }{$F_\pi(0)=1$.}
 
The inelastic threshold occurs, rigorously speaking, at $s=16m^2_\pi$. 
However, it is an experimental fact that inelasticity is negligible 
until the quasi-two~body channels $\omega\pi,\,a_1\pi\,\dots$ are open. 
In practice one can take
$$t_0\simeq 1\;\gev^2,$$
and fix the best value for $t_0$  empirically. 
It will be $t_0=1.1\,\gev^2$, and  the dependence 
of our results on  $t_0$ is very slight. 

The properties (i-iv) can be taken into account with the  
Omn\`es-Muskhelishvili method.\fnote{More details about the solution of the Omn\`es--Muskhelishvili 
equations can be found in the text of N.~I.~Muskhelishvili, {\sl Singular
 Integral Equations}, Nordhoof, 1958, and, applied to our present case, in ref.~6.} We
construct a  function $J(s)$ with the proper phase 
and asymptotic behaviour by defining 
$$J(s)=\ee^{1-\delta_1^1(t_0)/\pi}
\left(1-\dfrac{s}{t_0}\right)^{[1-\delta_1^1(t_0)/\pi]t_0/s}
\left(1-\dfrac{s}{t_0}\right)^{-1}
\exp\left\{\dfrac{s}{\pi}\int_{4m^2_\pi}^{t_0} \dd t\;
\dfrac{\delta_1^1(t)}{t(t-s)}\right\}.
\equn{(2.3a)}$$
We have written the dispersion relation with one subtraction to ensure 
that $J(0)=1$. 
The singular integral is understood to be calculated replacing $s\to s+\ii\epsilon$, 
$\epsilon>0$,   $\epsilon\to0$. We then define 
 the function $G$ by
$$F_\pi(s)=G(s)J(s),
\equn{(2.3b)}$$ 
and it follows from properties (i-ii) that $G(s)$ is analytic with
 only the exception of a cut from $t_0$ to infinity, 
as we have already extracted the correct phase 
below $s=t_0$. 

We can apply  the effective range theory to parametrize the phase  $\delta^1_1$. According to 
this, the function
$$\psi(t)\equiv \dfrac{2k^3}{t^{1/2}}\cot \delta_1^1(t),\quad k=\dfrac{\sqrt{t-4m^2_\pi}}{2}
\equn{(2.4a)}$$
is analytic in the variable $t$ except for two cuts: a cut from $-\infty$ to $0$, and 
a cut from $t=t_0$ to $+\infty$.
To profit from the analyticity properties of $\psi$ we will 
make a conformal transformation. We define 
$$w=
\dfrac{\sqrt{t}-\sqrt{t_0-t}}{\sqrt{t}+\sqrt{t_0-t}}.
$$
When $t$ runs the cuts, $w$ goes around the unit circle.
 We may therefore expand $\psi$ in a power series 
 convergent inside
the unit disc. In fact, because we know that the P wave resonates (which implies a zero 
of $\psi$) it is convenient to
expand not
$\psi$ itself, but the ratio   
$\psi(t)/(m^2_\rho-t)\equiv\hat{\psi}(t)$. 
Here $m_\rho$ is the mass of the rho resonance;  so we write,
$$\psi(t)=(m^2_\rho-t)\hat{\psi}(t)=(m^2_\rho-t)\left\{b_0+b_1w+
\cdots\right\}.
\equn{(2.4b)}$$

The P-wave, $I=1$ $\pi\pi$ scattering length, $a_1^1$, is related to $\psi$ by
$$a_1^1=\dfrac{1}{m_\pi\psi(4m^2_\pi)};
$$
experimentally, 
 $a_1^1\simeq (0.038\pm0.003) m^{-3}_\pi$, a condition that may be incorporated into the fit. 
 Note, however, that we do {\sl not} 
assume the values of $m_\rho,\,\gammav_\rho$. 
We only require that $\psi$ has a zero, and will let the fits 
fix its location and residue. 
It turns out that, to reproduce the width and scattering length, and to fit 
the pion form factor as well, only two $b_0$, $b_1$ 
 are needed in (2.4b). 

We now turn to
 the function 
$G(s)$. This function is analytic except for a cut from $s=t_0$ to $+\infty$. 
The conformal transformation 
$$z=\dfrac{\tfrac{1}{2}\sqrt{t_0}-\sqrt{t_0-s}}{\tfrac{1}{2}\sqrt{t_0}+\sqrt{t_0-s}}
$$
maps this cut plane into the unit circle. 
So we may write the 
expansion,
$$G(s)=1+c_1(z+1/3)+c_2 (z^2-1/9)+\cdots\,,
\equn{(2.5)}$$
which takes into account the condition $G(0)=1$ order by order.
We will only need two terms in the expansion, so we  have 
$c_1,\,c_2$ as free parameters.
This means that, altogether, we have the  five 
parameters,
$$m_\rho,\; b_0,\; b_1,\;c_1,\;c_2, 
$$
 to fit  158 experimental points.

One can then use the formulas just discussed and fit the experimental data 
on $F_\pi(s)$, after we have taken into account the $\omega-\rho$ interference (which includes the $\omega\to2\pi$
piece). 
This we do with the Gounnaris--Sakurai method. 
We write
$$F^{\rm all}_\pi(s)=F_\pi^{\rm bare}(s)\times
\dfrac{1+\sigma\dfrac{M^2_\omega}{M^2_\omega-s}}{1+\sigma},
\quad M_\omega=m_\omega-\ii\gammav_\omega/2
\equn{(2.6a)}$$
where $F_\pi^{\rm bare}$ is the form factor we would have in absence of 
$\omega-\rho$ interference and $m_\omega$ the (real) omega mass. 
 We take the values of $m_\omega$, $\gammav_\omega$ 
from the Particle Data Tables and find
$$|\sigma|=(18\pm1)\times10^{-4},\quad \arg\sigma=12\pm3\degrees.
\equn{(2.6b)}$$ 
The fit improves when using the 2002 Novosibirsk data;\ref{4}
there is better agreement with the data of Barkov et al.,\ref{4} 
in particular for the larger values of $s$.
The parameters of the fit are, however, similar  to what we found in 
ref.~6 using the old (1999) Novosibirsk data, as we shall see. 
 It should be noted that, in these data, 
 electromagnetic corrections have been  {\sl extracted}; 
we give the details of the procedure in \subsect~2.2 below.

The result of the contribution to the hadronic part of the muon anomaly is now, fitting the 
 $e^+e^-$ annihilation data,\ref{4,5}
$$\quad 10^{11}\times a^{(2)}_\mu(s\leq0.8\,{\gev}^2)=4\,707\pm21;\quad \chi^2/{\rm d.o.f.}=91/(114-7).
\equn{(2.7)}$$ 
The fact that the \chidof\ is substantially smaller than unity 
means that there is some room for displacement of the central value 
given in (2.7).

The result in (2.7) number may be improved, which  
 we will do in two steps. 
First of all, we remark that, 
because our  formulas for $F_\pi(s)$ are valid for spacelike as well as timelike
$s$, we can use information not only from $e^+e^-$ annihilation,\ref{4,5} 
but also from $\pi e$ scattering,\ref{7} 
which gives $F_\pi(s)$ for negative $s$. 
We record the results of two recent evaluations, made using this method. 
We have our evaluation here,
$$\eqalign{
[e^+e^-,\,\pi e]&:\cr
10^{11}\times a^{(2)}_\mu(s\leq0.8&\,{\gev}^2)=
\cases{
4\,750\pm19\quad[\hbox{TY; fixed norm.}];\;  \dfrac{\chi^2}{\rm d.o.f.}=
\dfrac{172}{159-7}\vphantom{\Bigg|}\cr
4\,715\pm20\pm25\quad[\hbox{TY; float. norm.}];\;  \dfrac{\chi^2}{\rm d.o.f.}=
\dfrac{134}{161-10}.
\cr}\cr}
\equn{(2.8a)}$$
The errors in the first number here do not include systematic errors in $e\pi$ scattering, 
which is 
the reason for the largish \chidof\ 
In the second set, the first error is the statistical error, the second the systematic one. 
We  take into account the systematic errors, both in $e^+e^-$ data\ref{4} and in the 
$e\pi$ data\ref{7} 
 as in ref.~6, by including a factor $1+\epsilon$ in the normalization of each set of data,
and allowing $\epsilon$ to float. 
We find 
$$\eqalign{
\epsilon_{e\pi\,[{\rm ref.~7}]}=(1.3\pm0.2)\%,&\,\quad \epsilon_{e^+e^-\,[{\rm Akhmetshin,\;ref. 4}]}=
(-0.4\pm0.6)\%,\cr
\epsilon_{e^+e^-\,[{\rm OLYA;\; Barkov,\;ref. 4}]}=&\,(0.5\pm0.9)\%;\cr
}
\equn{(2.8b)}$$
the fit is depicted in \fig~2.  
The only $\epsilon$ that is not compatible with zero 
is $\epsilon_{e\pi}$. In fact, we left its value arbitrary; that 
the result we found is 
consistent with the value quoted in ref.~7 itself, $\epsilon_{e\pi}\simeq1\%$, 
is a nontrivial test of the quality of data in this reference 
(in particular, of their estimate of systematic errors) and of the 
consistency of our fitting procedure.

Then, we have a recent result by Colangelo and collaborators,\ref{8}
$$[e^+e^-,\,\pi e]:\quad 10^{11}\times a^{(2)}_\mu(s\leq0.8\,{\gev}^2)=
4\,679\pm30\quad[\hbox{Colangelo, ref.~8}].
\equn{(2.9)}$$ 
The  evaluation of Colangelo and collaborators, 
Eq.~(2.9), includes  the four-pion cut, and  
 imposes the P wave phase shift as given 
by chiral-dispersive evaluations.\ref{9} 
This last feature, however, makes the Colangelo result 
vulnerable to possible defects of the chiral-dispersive evaluation, such as those 
discussed in ref.~10.

\topinsert{
\setbox0=\vbox{{\epsfxsize 12.2truecm\epsfbox{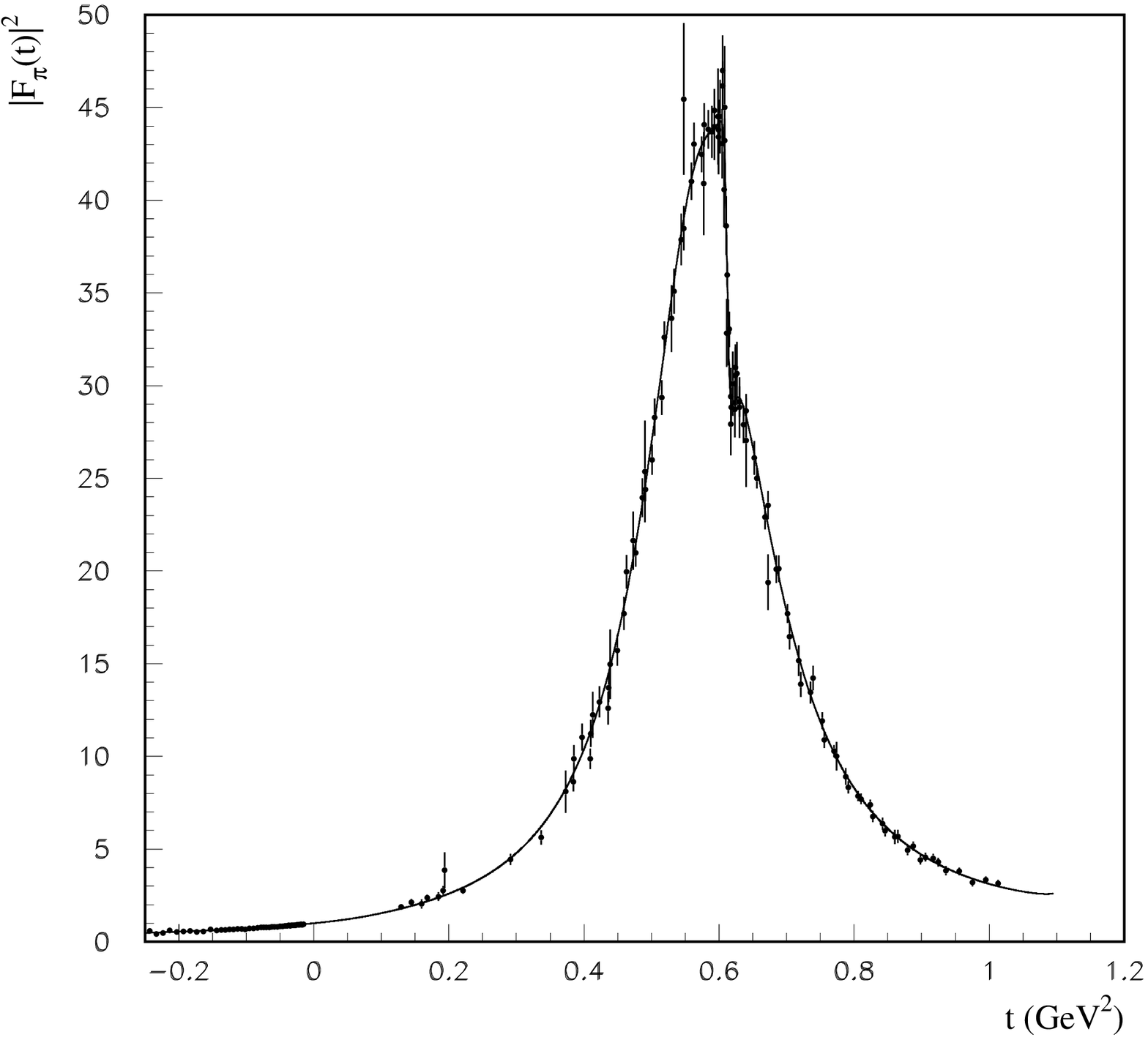}}} 
\setbox6=\vbox{\hsize 10.4truecm\captiontype\figurasc{Figure 2. }{The fit 
to the pion form factor data in 
the timelike and spacelike regions;  the more recent (2002) Novosibirsk data are included.
}} 
\centerline{\box0}
\medskip\centerline{\box6}
\medskip}\endinsert

A  way to improve the precision of (2.7) is to include information on 
$\tau$ decay\ref{11} in the fit. 
We consider the weak vector current correlator, related to $\tau$ decay,
and expand it  as
$$\eqalign{
\Piv^V_{\mu\nu}=&\,
\ii\int\dd^4x\,\ee^{\ii p\cdot x}\langle0|{\rm T}V^+_\mu(x)V_\nu(0)|0\rangle=
\left(-p^2g_{\mu\nu}+p_\mu p_\nu\right)\Piv^V(t)+p_\mu p_\nu \piv^{S}(t);\cr
v_1\equiv&\,2\pi\imag \Piv^V,\quad t=p^2.\cr
}
\equn{(2.10a)}$$  
We can then write 
$$v_1=\tfrac{1}{12}
\left\{\left[1-\dfrac{(m_{\pi^+}-m_{\pi^0})^2}{t}\right]
\left[1-\dfrac{(m_{\pi^+}+m_{\pi^0})^2}{t}\right] \right\}^{3/2}|F^{(\tau)}_\pi(t)|^2.
\equn{(2.10b)}$$
To compare with the experimentally measured quantity, 
which involves all of $\imag \piv^V_{\mu\nu}$, we have to 
neglect the scalar component $\piv^S$, which is proportional to $(m_d-m_u)^2$, 
and thus very small. Moreover, 
$F^{(\tau)}_\pi$ only equals $F_\pi$ in the limit of exact isospin invariance. 
We have also complications due to radiative corrections, that we will discuss 
in \subsect~2.3.

One can take into account isospin breaking effects, at least partially,
by realizing that the mass and widths of the $\rho^0$ and $\rho^\pm$ can be different.\ref{6} 
One finds,
$$\eqalign{
[e^+e^-,\,\pi e,\,\tau]&:\cr
 10^{11}\times a^{(2)}_\mu(s\leq&0.8\,{\gev}^2)=\cases{
4\,793\pm17\,({\rm St.}),\quad \dfrac{\chi^2}{\rm d.o.f.}=\dfrac{283}{241-9};\;[\hbox{TY}]\cr
4\,798\pm17\,({\rm St.})\pm25\,({\rm Sys.}),\; \dfrac{\chi^2}{\rm d.o.f.}=\dfrac{245}{244-13};\;[\hbox{TY,
Sys.}]\cr }\cr}
\equn{(2.11a)}$$
depending on whether or not one takes into account systematic normalization errors. 
We unify the normalization of the tau decay 
data taking into account the relevant branching ratios,
as  
given in the Particle Data Tables.\fnote{When 
quoting the PDT we refer to the 2002 edition,  K.~Hagiwara  et al., {\sl Phys. Rev.} 
{\bf D66}, 010001 (2002).}

In the second set of numbers in (2.11), we have, as in (2.8), taken into account systematic errors 
by allowing  
floating  normalization by a factor $1+\epsilon$ of the various data sets. 
We find,
$$\eqalign{
\epsilon_{e\pi\,[{\rm ref.~7}]}=(1.0\pm0.2)\%\quad&\,[1\%],\qquad 
\epsilon_{e^+e^-\,[{\rm Akhmetshin,\;ref. 4}]}=(0.4\pm0.5)\%\quad[0.6\%],\cr
\epsilon_{\tau}=(-1.4\pm0.5)\%\quad&\,[0.7\%],
\quad\epsilon_{e^+e^-\,[{\rm OLYA;\;Barkov,\;ref.~4}]}=(2.0\pm0.8)\%\quad[4\%]  .\cr
}\equn{(2.11b)}$$
In square brackets we give the estimate of the normalization errors 
as given by the experimental groups themselves, 
except for $\tau$ decay, where the number  $0.7\%$ is taken 
from the  Particle Data Tables. 
The only case where the value we find for $\epsilon$ exceeds the expectations is for 
$\tau$ decay data, although the difference is very small, 
$0.7\pm0.5\%$. 
We will discuss this again in \sect~2.3.

The best number in (2.11a), of course, is the second, $4\,798\pm31$. 
The fact that the \chidof\ is still a bit larger than unity 
can be traced partly 
to the size of $\epsilon_\tau$, 
that we discuss in 
\subsect~2.3, and partly to a discrepancy between the tau decay data of OPAL and of
ALEPH and CLEO. This is seen very clearly if we give the individual 
values for the ratio of $\chi^2$ to number of experimental points of 
the various sets of data, 
which we do for the second fit in (2.11a), i.e., including systematic errors:
$$\matrix{
e\pi\quad[{\rm NA7,\; ref.~7}]:\quad&42/45\cr
e^+e^-\quad\quad\;[{\rm ref.~4}]:\quad&108/113\cr
\tau\; {\rm decay}\;[{\rm Aleph,\;ref.~11}]:\quad&19/21\cr
\tau\; {\rm decay}\;[{\rm Cleo,\;ref.~11}]:\quad&32/30\cr
\tau\; {\rm decay}\;[{\rm Opal,\;ref.~11}]:\quad&40/31.\cr
}\equn{(2.11c)}$$
A remarkable feature of (2.11c) is that it shows that 
including information from tau decay in the fit does not spoil the 
quality of the fit of the pure $e^+e^-$ data.

\booksubsection{2.2. Radiative corrections for the $e^+e^-$ case}

\noindent
We next devote a few words to discuss electromagnetic  radiative corrections, 
a subject of crucial importance given the precision of 
the more recent data, but not always very clear in the existing 
literature.

\topinsert
{
\setbox0=\vbox{\hsize8.4truecm{\epsfxsize 7.truecm\epsfbox{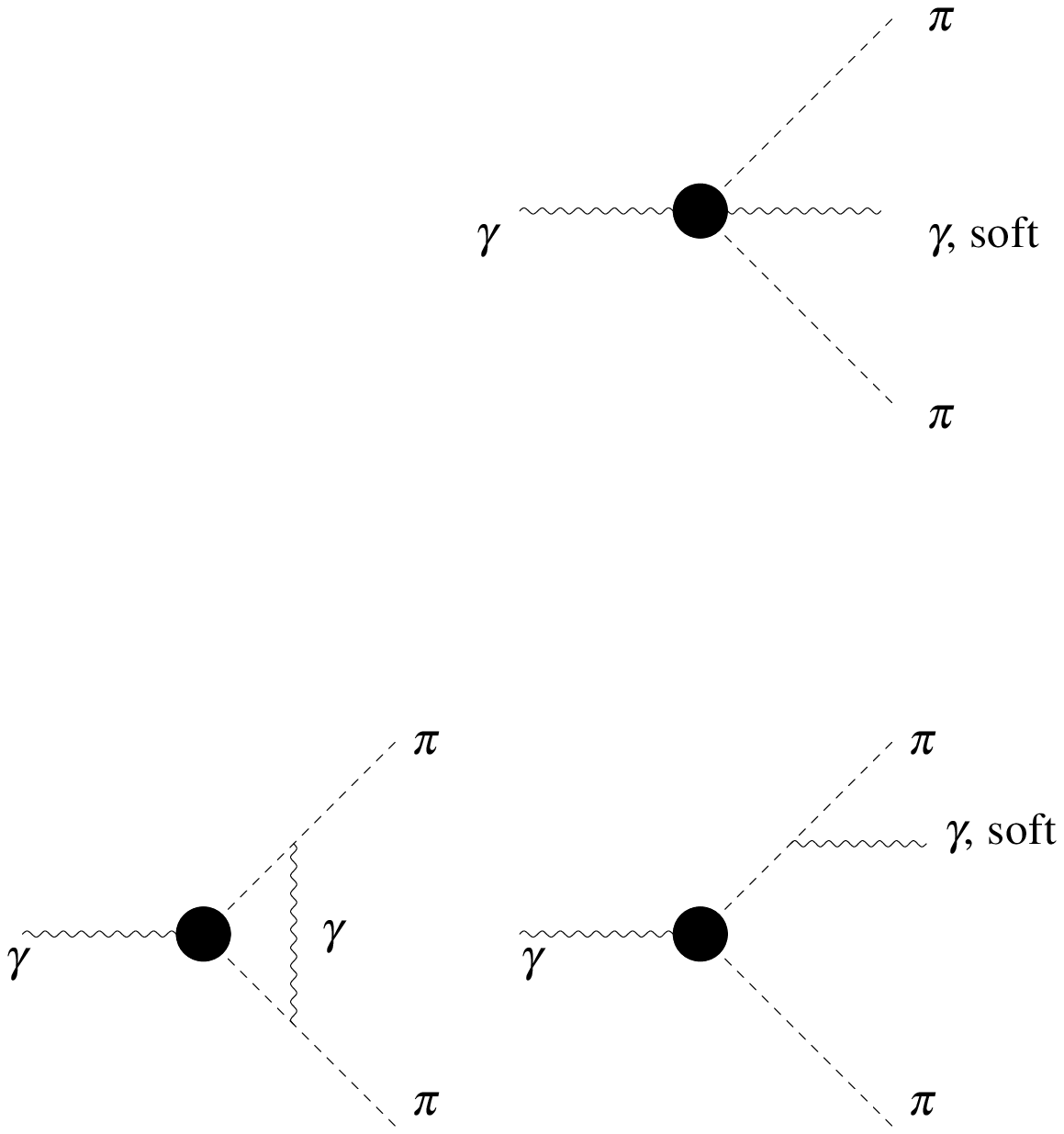}}\hfil}
\setbox1=\vbox{\epsfxsize 4.2truecm\epsfbox{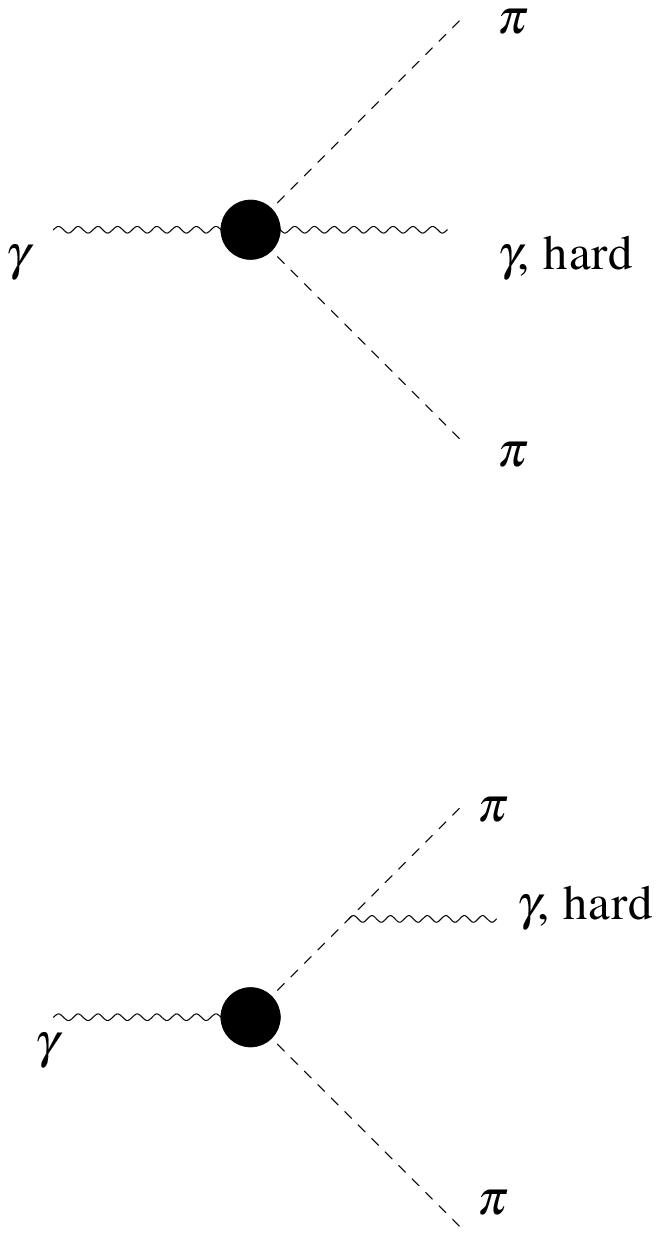}}
\setbox2=\vbox{\hsize 10.truecm\hfil ({\sc A})\kern16em({\sc B})\hfil} 
\setbox6=\vbox{\hsize 10.truecm\captiontype\figurasc{Figure 3. }{Diagrams 
subtracted for evaluating the pion form factor contribution in 
$a_\mu^{(2)}(s\leq0.8\,\gev^2)$, 
but included in the $O(\alpha^3)$ contribution to $a_\mu({\rm Hadr.})$. 
The blob represents the pion form factor, to zero order in 
electroweak interactions, as defined in Eq.~(2.12).}\hb
\vskip.1cm} 
\line{\kern-1.4truecm\box0\hfil\box1}
\centerline{\box2}
\bigskip
\centerline{\box 6}
\medskip
}
\endinsert

There are, in fact, two separate questions here. 
First, we have the radiative corrections to the hadronic part of the 
photon vacuum polarization, which affect Fig.~1 by adding photon corrections 
(depicted below, in Fig.~6).  
These give corrections to $a_\mu$ of order $\alpha^3$ and will be considered 
later, in Sect.~3. 
Secondly, we have the matter of the 
radiative corrections that have to be taken into account when extracting the 
pion form factor from experimental data. These come about for the following reason: 
the form factor that verifies the analyticity
 and unitarity properties necessary to
 carry out our analysis in \subsect~2.1 is defined
by 
$$\langle p|J_\mu(0)|p'\rangle\Big|_{\rm electroweak\;int.\equiv0}=
(2\pi)^{-3}(p-p')_\mu F_\pi((p-p')^2),
\equn{(2.12)}$$
i.e., only strong interactions are taken into account for the expectation
 value of the electromagnetic current, 
$J_\mu$.

This quantity, $F_\pi$, has thus to be extracted from the experimentally 
measured cross sections for 
$e^+e^-\to\pi^+\pi^-$, which include all sorts of radiative corrections. 
To first (relative) order in $\alpha$,
these are the following: I) Corrections to the
$e^+e^-\gamma$ vertex,  or photon radiation by the incoming
$e^+,\,e^-$. 
These are pure QED effects readily calculated and taken into account as a matter of 
course in experimental analyses. II) Vacuum polarization corrections 
to the photon propagator. These are known in terms of ${\rm Im}\, \Piv$, 
and are also be subtracted easily. 
They are explicitly taken into account in 
the second paper in ref.~4. III) Corrections to the
$\pi^+\pi^-\gamma$ vertex,  or photon radiation by the outgoing pions 
(Fig.~3). We now say a few words about the last. 

Radiation of hard photons by the outgoing pions, as in Fig.~3B, is
excluded by the  experimental cuts applied when analyzing 
$e^+e^-\to\pi^+\pi^-$ scattering, 
 which require the angle
between  the momenta of $\pi^+,\,\pi^-$ to be close to 180$^{\rm o}$. 
One is thus left with the soft photon radiation and vertex correction 
shown in Fig.~3A. They can be calculated under the 
assumption that one can factorize the pion form factor and, 
given the actual values of the experimental cuts applied to the momenta of the $\pi^+,\,\pi^-$, 
this correction turns out to be very small. 
In this approximation, the corrections have been evaluated long ago and 
 are, fortunately, explicitly given and extracted in the 2002 
version of the Novosibirsk data (ref.~4).

From this analysis 
it follows that, for the $e^+e^-\to\pi^+\pi^-$ case, one can subtract all corrections 
and really obtain $F_\pi$, as defined in Eq.~(2.12), from data 
with an error that is only  of order $\alpha^2$. 
This is the quantity to which we can apply our theoretical analysis, 
as we did at the
beginning of 
the present Section.

A problem with the evaluation of the radiative corrections here, however, is that, 
as noted,  
one is using a model with elementary pions, 
in which the form factor is included by hand (factorized). 
This may cause errors (for example, 
due to rescattering of the pions or dependence of the form factor on the 
off-shell pion mass), whose size (likely small, since the correction itself is small) 
we will estimate when discussing the case of tau decay in next Subsection.

We should here note that the radiative corrections to the
$\pi^+\pi^-\gamma$ vertex,  or photon radiation by the outgoing pions
will have to be considered again when we consider contributions to $a_\mu$
due to photon vacuum polarization  of $O(\alpha^3)$.
This we will do in Sect.~3.

\booksubsection{2.3. Comment on combining results using $e^+e^-\to\pi\pi$ and 
$\tau\to\nu\pi\pi$, and on radiative corrections for $\tau$ decay}

\noindent
In some recent papers much ado is made about the difference 
in the pion form factor, in the region near and above the 
rho resonance, depending whether it its extracted from $e^+e^-\to\pi\pi$ or
from 
$\tau\to\nu\pi\pi$. 
In fact, this difference is expected. 
To begin with, $|F(s)|^2$ grows almost by a factor 
 50 around the rho.\fnote{There is also a significant difference 
between data from $e^+e^-$ and tau decay for $0.8\,\gev^2<s<1.1\,\gev^2$ 
which, however, affects very little the result for  $a_\mu(s\leq0.8\,{\gev}^2)$.} 
Thus, even a small difference between the masses and widths of $\rho^\pm$ and $\rho^0$ will 
imply a large difference in the form factors. 
This matter was studied carefully in ref.~6, 
which discussion we summarize and update now. 

The values for the anomaly in the rho region that one obtains depending 
what one fits, is, with only statistical errors,
$$10^{11}\times a^{(2)}_\mu(s\leq0.8\,{\gev}^2)=\cases{
4\,707\pm21\quad[\;{\rm from}\;e^+e^-\to\pi\pi\;{\rm only}]\cr
4\,820\pm11\quad[\;{\rm from}\;\tau\to\nu\pi\pi\;{\rm only}].\cr
}
\equn{(2.13)}$$
For the tau, we have $\chidof=87/(83-5)$. 
The two numbers are well outside each other error bars and it is,
 therefore, dangerous to combine them 
in a direct manner. 
What, however, one can do, is to fit simultaneously  $e^+e^-\to\pi\pi$ and 
$\tau\to\nu\pi\pi$ data allowing for different values of the parameters
 $m_\rho$, $b_0$, $b_1$, 
(and therefore, also different widths\fnote{We 
take, however, equal P wave  scattering 
lengths for $\pi^+\pi^-$ and $\pi^0\pi^\pm$. 
We have checked that the influence of this is negligible.}) but
 with the {\sl same}  Omn\`es--Muskhelishvili function
$G(s)$ in 
\equn{(2.3b)}. 
As discussed in ref.~6, we expect isospin breaking effects to be small for $G(s)$ 
since its imaginary part is different from 
zero only for $s>1.1\,\gev^2$. 

Of course, when calculating $a_\mu(s\leq0.8\,{\gev}^2)$ 
one uses the parameters $m_\rho$, $b_0$, $b_1$ determined from the fit to 
$e^+e^-\to\pi\pi$, even if we 
use tau decay data to help fix $G$:
 we should perhaps emphasize that 
the result  reported in (2.11) is {\sl not} an 
average of $e^+e^-$ and $\tau$ results, but an evaluation of 
$e^+e^-\to\pi\pi$, using information on 
$G(s)$ from tau decay. 

This use of different masses and widths for $\rho^\pm$, $\rho^0$
 is not sufficient to remove the 
discrepancies between the form factors obtained from $e^+e^-$ and $\tau$ decay.
 A  reason for at least part of the remaining discrepancy  
 is that the experimental number given in ref.~11 for $\tau$ decay includes also 
the radiative decay; that is to say, one does not measure the quantity  
$\gammav(\tau\to\nu\pi\pi)$ but, in fact,  
$\gammav(\tau\to\nu\pi\pi)+\gammav(\tau\to\nu\pi\pi+\gamma)+\hbox{higher orders}$. 
So we should discuss radiative corrections also for  tau decay.

\topinsert{
\setbox0=\vbox{\hsize12.4truecm{\epsfxsize 11.truecm\epsfbox{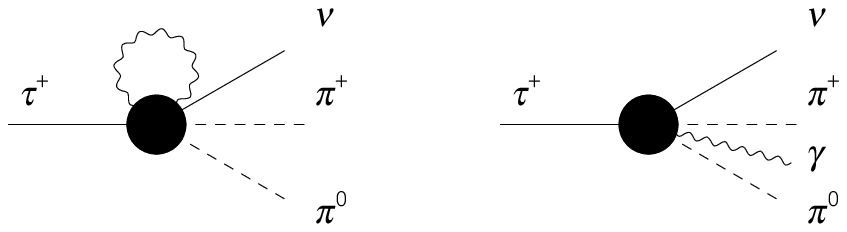}}} 
\setbox6=\vbox{\hsize 10truecm\captiontype\figurasc{Figure 4. }{Radiative 
corrections to $\tau\to\nu\pi\pi$ decay.}} 
\centerline{\box0}
\centerline{\box6}
\medskip
}\endinsert

First of all, we must make more explicit what we used in our fits. 
We take the experimental numbers for 
the decay $\tau\to\nu\pi\pi$ including an eventual photon. 
The corresponding width we denote by 
$\gammav_{\rm exp.}(\tau\to\nu\pi^\pm\pi^0(\gamma))$, and we divide this by the 
experimental decay rate for $\tau\to\nu_\tau e\nu_e$, plus eventual gammas, 
$\gammav_{\rm exp.}(\tau\to\nu_\tau e\nu_e(\gamma))$. 
However, and as in the case of $e^+e^-$ annihilation, we would like to have the 
quantity $F^{(\tau)}_\pi$ {\sl with electroweak interactions set to zero}. 
Therefore, we should correct the ratio
$$\dfrac{\gammav_{\rm exp.}(\tau\to\nu\pi^\pm\pi^0(\gamma))}
{\gammav_{\rm exp.}(\tau\to\nu_\tau e\nu_e(\gamma))}$$
to obtain the quantity
$$
\dfrac{\gammav^{(0)}(\tau\to\nu\pi^\pm\pi^0)}
{\gammav^{(0)}(\tau\to\nu_\tau e\nu_e)}$$
where the $\gammav^{(0)}$ are evaluated to lowest order in electroweak interactions.

The corrections necessary to do this are as follows. 
We first have radiative corrections to the leptonic width, 
 that give
$$\gammav_{\rm exp.}(\tau\to\nu_\tau e\nu_e(\gamma))=
(1+\delta_e)\,\gammav^{(0)}(\tau\to\nu\pi^\pm\pi^0),\quad 
\delta_e\simeq\left(\tfrac{25}{4}-\pi^2\right)\dfrac{\alpha}{2\pi}
\simeq-0.004.$$ 
This correction was incorporated in the analysis of \subsect~2.1.

Then we have the corrections to the hadronic width.
We here have corrections similar to those in \fig~3{\sc B}; and 
diagrams similar to those in \fig~3{\sc A}: see \fig~4. 
We start with the first, that is to say, the sum of loop corrections
 (plus radiation of 
a soft photon). 
As far as we know, this 
has not been calculated exactly. 
However, we may expect this to be dominated by the short distance piece, since 
this last contains a large logarithm, $\log M_Z/m_\tau$. 
This gives the correction 
$$1+\dfrac{2\alpha}{\pi}\log\dfrac{M_Z}{m_\tau}\simeq1.019
\equn{(2.14)}$$
as  has  been known from a long time.\ref{12} 
This piece we have also extracted  
in our analysis above.
After so doing, there still remains a piece of the loop correction and 
a correction due to soft photon radiation. 
Since we have extracted the large $\log M_Z/m_\tau$ piece, we expect
 this to be comparable to the 
like piece in $e^+e^-\to\pi\pi$ case, and thus small.
 
It remains to correct for the ratio
$$1+\delta_\gamma\equiv
\dfrac{\gammav(\tau\to\nu\pi\pi)+\gammav(\tau\to\nu\pi\pi+\gamma)}
{\gammav^{(0)}(\tau\to\nu\pi\pi)},
\equn{(2.15)}$$
with $\gamma$ a hard photon, i.e., a photon with energy larger than a given $E_0$.
This, again, is not known, but one can approximate it by the infrared logarithmic piece, 
which gives an average correction
$$\delta_\gamma\simeq \dfrac{\alpha}{\pi}\log\dfrac{m_\tau-\bar{M}_{\pi\pi}}{E_0}\simeq0.008,
\equn{(2.16)}$$
the last for $20\,\mev\leq E_0\leq80\,\mev$, 
and we have neglected terms of $O(m_\pi/m_\tau)$.
$\bar{M}_{\pi\pi}$ is the average invariant mass of the two pions, that we take equal to
$m_\rho$.  Since this correction plus the remainder (after extracting the 
logarithm) of the loop correction are not known exactly, we have {\sl not} 
included the correction (2.16) in our evaluations above: 
(2.16) will be part of the normalization factor 
$\epsilon_\tau$.
 
When we allowed variations of the normalization for tau decay data,\ref{11}  
multiplying their numbers by $1+\epsilon$, and letting $\epsilon$  float, 
we found the excellent fit reported in (2.11) with 
$\epsilon_\tau=-1.4\%$. 
It unfortunately is not possible to understand all of this as due only to the 
neglect of radiative corrections; 
we have verified that, including $\delta_\gamma$ as given in 
(2.16) in the fit only changes $\epsilon_\tau$ to $-1.2\%$. 
We have to admit that there is a residual discrepancy with the value 
for the error in normalization, $|\epsilon_\tau|=0.7\%$, given 
in the Particle Data Tables; but, the difference is small 
and, indeed,  we  get a 
 fit to all experimental data with a $\chi^2$ per experimental point 
which is essentially  unity. 

In spite of this, it is clear that here we have not fully
 determined radiative corrections;  only the logarithms $\log M_Z$, 
$\log E_0$ are exact in Eqs.~(2.14,~16). 
Moreover, the function $G(s)$ will not be exactly invariant under isospin and, 
finally, a correction due to the fact that $m_u\neq m_d$, 
although likely very small, also exists beyond the rho. 
We must thus conclude that our partial ignorance of isospin violations, 
very likely dominated by
radiative corrections, 
implies a possible shift of the central value of the anomaly. 
A conservative estimate for these effects would be the difference between 
the two values obtained leaving the 
$\tau$ decay normalization fixed, and the same allowing it to float. 
This gives the number 
$$\deltav_{\rm rad}\left[10^{11}\times a^{(2)}_\mu(s\leq0.8\,{\gev}^2)\right]
\simeq10.
\equn{(2.17)}$$
We will accept the same error for the process 
$e^+e^-\to\pi\pi$, 
although it is probably smaller here.

\booksubsection{2.4. The region with $s>0.8\,\gev^2$}

\noindent
For low energies (say, $s<2\,\gev^2$), and near quark thresholds,
 there is no alternative to using experimental
data.\ref{12} 
Between $0.8\,\gev^2$ and $1.2\,\gev^2$ 
we fit the experimental data\ref{4} for the 
$\pi^+\pi^-$ channel; 
its contribution is $(229\pm3\pm3)\times10^{-11}$. 
For other final states ($\bar{K}K$, $3\pi$, $4\pi$,\tdots) we use
 the  $e^+e^-$ data of refs.~5 and 13, 
with the methods of ref.~6.
For higher energies, $s>2\,\gev^2$, and   away from quark 
thresholds,  we can use QCD formulas,\ref{14} taking into account the more 
recent values of the masses of the quarks as 
well as the strong coupling, $\alpha_s$.
For the QCD calculations we take the following approximation: for $n_f$ massless quark 
flavours, 
with charges $Q_f$, we write
$$\eqalign{R^{(0)}(s)=&\;3\sum_fQ_f^2\Bigg\{1+\dfrac{\alpha_s}{\pi}+
(1.986-0.115n_f)\left(\dfrac{\alpha_s}{\pi}\right)^2\cr
+&\;\Big[-6.64-1.20n_f-0.005n_f^2-1.240\dfrac{(\sum_f Q_f)^2}{3(\sum_f Q_f^2)}\Big]
\left(\dfrac{\alpha_s}{\pi}\right)^3\Bigg\}.\cr}
$$
To this one adds mass and nonperturbative corrections. 
We take into account the $O(m^{2})$ effect for $s,\,c,\,b$ quarks with 
running masses $\bar{m}_i(s)$, which correct $R^{(0)}$ by the amount
$$-3\sum_{i=s,c,b}Q_i^2\bar{m}^2_i(s)\left\{6+28\dfrac{\alpha_s}{\pi}
+(294.8-12.3n_f)\left(\dfrac{\alpha_s}{\pi}\right)^2\right\}s^{-1}.
$$
The details may be seen in refs.~6,~15. 

Adding all the contributions, one has
$$10^{11}\times a^{(2)}_\mu(s\geq0.8\,{\gev}^2)=2\,134\pm35.
\equn{(2.19)}$$
This number is almost the same that may be found in ref.~6, using the 1999 Novosibirsk 
data; including 
the new data has very little influence in this region.

\booksection{3. The hadronic contributions to $a_\mu$. II: the $O(\alpha^3)$ pieces}

\noindent
A contribution in a class by itself is the hadronic light by light one, 
that we label ``$\ell\times \ell$". 
So we split
$$a(\hbox{Hadronic, $O(\alpha^3)$})=a(\hbox{`One blob' hadronic, $O(\alpha^3)$})+
a(\hbox{$\ell\times \ell$}).\equn{(3.1)}$$

\topinsert
{
\setbox0=\vbox{\hsize8.4truecm{\epsfxsize 7.truecm\epsfbox{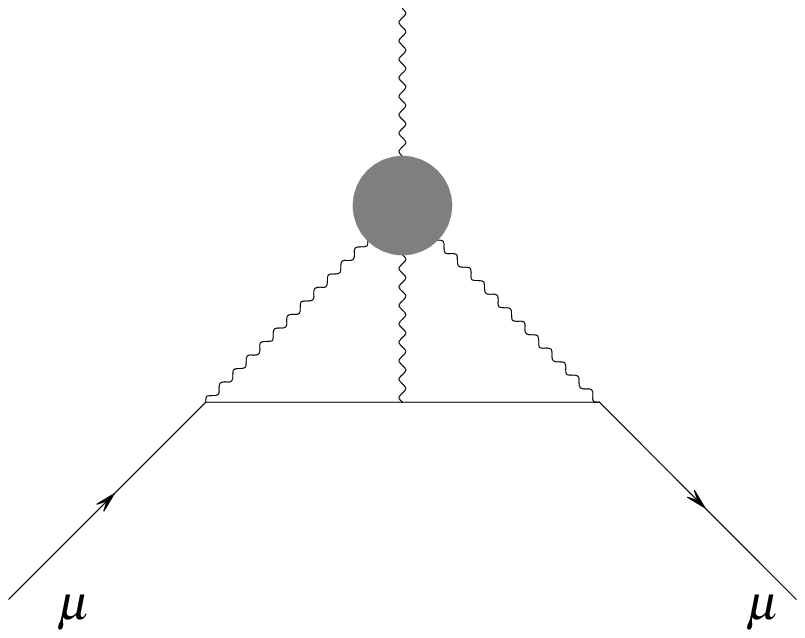}}} 
\setbox6=\vbox{\hsize 4.5truecm\captiontype\figurasc{Figure 5. }{A typical 
diagram for the  
hadronic light by light contributions to the muon magnetic moment.}\hb
\vskip.1cm} 
\medskip
\line{\box0\hfil\box6}
\medskip
}
\endinsert

We will start by considering the last, given diagrammatically by  graphs 
like that of \fig~5.
This can be evaluated  only using {\sl models}. 
One can make a chiral model calculation (essentially, replacing the hadronic 
blob by the lightest hadronic state, a $\pi^0$),  or one can use a constituent
quark model  in which we replace the blob in \fig~5 by a quark loop. 
For the chiral model calculation we have to introduce a cut-off, 
since the $\pi^0$ contribution diverges for large virtuality of the 
photon lines. 
The result depends on the cut-off (for the chiral calculation) or on the constituent mass chosen 
for the quarks.
After the correction of a sign error in the 
evaluations of ref.~16 (see ref.~17) we find 
$$10^{11}\times a(\hbox{$\ell\times \ell$})=
86\pm25\quad \hbox{[Chiral calculation]}.
\equn{(3.2a)}$$ 
Earlier calculations with the $\pi^0$ model, using VMD to cure its divergence, gave 
(HKS, ref.~16)
$$10^{11}\times a(\hbox{$\ell\times \ell$})=
52\pm20\quad \hbox{[$\pi^0$ pole (HKS)]}.
\equn{(3.2b)}$$
 
One could also take the estimate  of the $\pi^0$ pole  from 
Hayakawa, Kinoshita and Sanda\ref{16} {\sl and} add 
the constituent quark loop, in which case we get
$$10^{11}\times a(\hbox{$\ell\times \ell$})=
98\pm22\quad \hbox{[Quark const. model+ pion pole]}. 
\equn{(3.2c)}$$
 One expects the $\pi^0$-dominated calculation to be valid for small 
values of the virtual photon momenta, 
and  the constituent model to hold for large values of the same.
Thus, almost half of the contribution to $a(\hbox{$\ell\times \ell$})$ 
in the chiral calculation comes from a region of momenta above $0.5$ GeV, 
where the chiral perturbation theory starts to fail,
while for this range of energies, and 
at least for the imaginary part of 
(diagonal) light by light scattering,  
 the quark model reproduces reasonably well the experimental data, 
as measured in photon-photon scattering. In view of this, we
 will take here the figure 
$$10^{11}\times a(\hbox{$\ell\times \ell$})= 92\pm30, 
\equn{(3.3)}$$  
but will refrain from combining this error with the others.

\topinsert
{
\setbox0=\vbox{\hsize8.4truecm{\epsfxsize 7.truecm\epsfbox{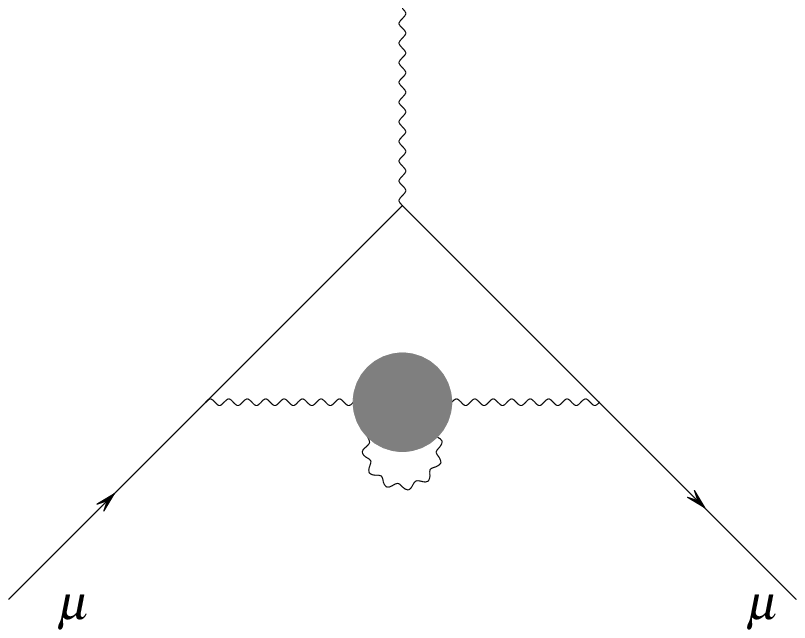}}} 
\setbox6=\vbox{\hsize 4.5truecm\captiontype\figurasc{Figure 6. }{The
$O(\alpha^3)$ hadronic
 correction\hb $a({\rm h.v.p.},\;\gamma)$.}\hb
\vskip.1cm} 
\medskip
\line{\box0\hfil\box 6}
\medskip
}\endinsert

We next turn to the   
$a(\hbox{`One blob' hadronic, $O(\alpha^3)$})$ corrections, which  
 are obtained by attaching a photon or fermion loop to the various lines in \fig~1. 
They can be further split into two pieces: the piece where both ends 
of the photon line are  attached  to the 
hadron blob, $a({\rm h.v.p.},\;\gamma)$, {\sl hadronic vacuum polarization} 
corrections, shown in \fig~6, and the rest. So
we write,
$$a(\hbox{`One blob' hadronic, $O(\alpha^3)$})=a({\rm h.v.p.},\;\gamma)+
a(\hbox{`One blob' hadronic, rest}).
\equn{(3.4)}$$
The last can be evaluated \ref{18} in terms of 
the hadronic contributions to the photon vacuum polarization, finding
$$10^{11}\times a(\hbox{`One blob' hadronic, rest})=-101\pm6.
\equn{(3.5)}$$
This result has been checked independently recently,  by the Marseilles group (S.~Friot, 
D.~Greynat; E.~de~Rafael, 
private communication) and in ref.~21.

The only contribution that requires further discussion is 
that depicted in \fig~6, $a({\rm h.v.p.},\;\gamma)$. 
In principle, this contribution can be evaluated straightforwardly 
by a generalization of the  method used to evaluate the $O(\alpha^2)$ contributions. 
We can write
$$a^{(2)}({\rm Hadr.})+a({\rm h.v.p.},\;\gamma)=\int_{4m^2_\pi}^\infty \dd t\,K(t)R^{(2)}(t),
\equn{(3.6)}$$
where 
$$R^{(2)}(t)=
\dfrac{\sigma^{(0)}(e^+e^-\to {\rm hadrons})+\sigma^{(2)}(e^+e^-\to {\rm hadrons})+
\sigma^{(0)}(e^+e^-\to {\rm hadrons};\,\gamma)}{\sigma^{(0)}(e^+e^-\to\mu^+\mu^-)}.
$$
The notation means that we evaluate the hadron annihilation cross section to second order in 
$\alpha$, and we add to it the first order annihilation into hadrons plus a photon. 
For energy  large enough  
this can be calculated with the parton model, which 
leads to a (very small) correction, $(2\pm1)\times10^{-11}$.

Then comes the  contribution of small 
momenta. 
We start by discussing the process involving two pions.
We calculate the corresponding piece by adding the contribution  
of the diagrams in \fig~3, as given in the 2002 paper by Akhmetshin et al.\ref{4}
In this way, we find   
$$10^{11}\times a({\rm h.v.p.},\;\pi^+\pi^-\gamma)=47.6\pm 0.3.
\equn{(3.7)}$$
The number is very close to that obtained in ref.~6 ($46\pm9$) but the errors have decreased 
drastically.\fnote{In ref.~6 only the 
radiation of hard photons was included, evaluated using the results of 
ref.~19. 
The fact that this is so similar to the full result justifies the (expected) 
smallness of soft photon plus vertex correction.}
A similar analysis ought  to be made, in principle, for other 
radiative intermediate states like $3\pi+\gamma$ and $\bar{K}K+\gamma$, 
which can be estimated in terms of the corresponding decays of 
the $\omega$ and $\phi$, but they give a contribution 
below the $10^{-11}$ level and we neglect them.
 The contribution from $\pi^0\pi^0\gamma$, $(2.0\pm0.3)\times10^{-11}$,
 is taken from
ref.~7.

The  lowest energy contributions to $\sigma^{(0)}(e^+e^-\to\,{\rm hadrons};\gamma)$ 
are those of the intermediate states $\pi^0\gamma$ and $\eta\gamma$, 
\fig~7. These contributions were evaluated in detail in ref.~6; we have 
$$\eqalign{
10^{11}\times a({\rm h.v.p.},\;\pi^0\gamma)=&\;37\pm3\cr
10^{11}\times a({\rm h.v.p.},\;\eta\gamma)=&\;6\pm2\cr
}
\equn{(3.8)}$$

\topinsert 
{
\setbox0=\vbox{\hsize8.2truecm{\epsfxsize 6.7truecm\epsfbox{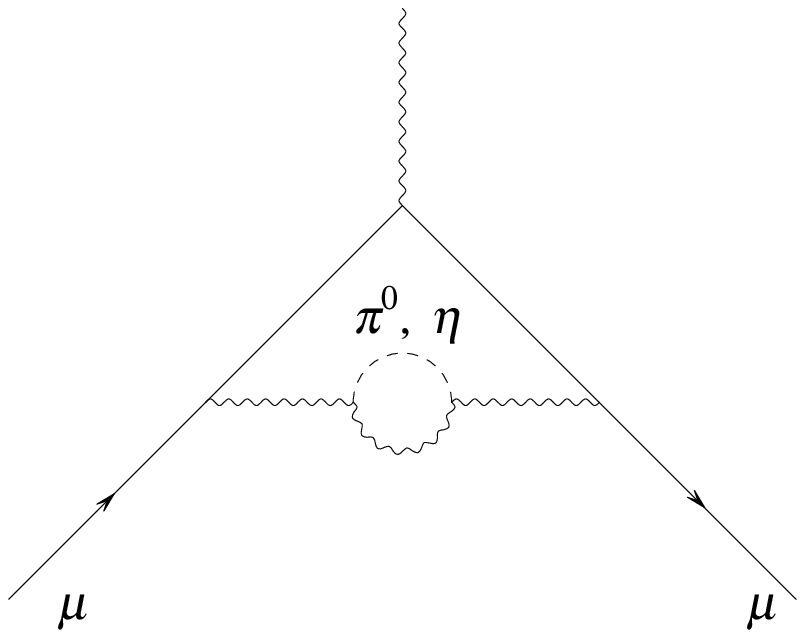}}} 
\setbox6=\vbox{\hsize 4.2truecm\captiontype\figurasc{Figure 7. }{The 
$\pi^0\gamma$, $\eta\gamma$  contributions to $a({\rm h.v.p.},\;\gamma)$.}\hb
\vskip.1cm} 
\medskip
\line{\box0\hfil\box6}
\medskip
}
\endinsert
 Collecting all of this, we  get the total 
  effect of the states $hadrons+\gamma$, 
including the loop correction,
$$10^{11}\times a({\rm h.v.p.},\;\gamma)=95\pm6.
\equn{(3.9)}$$

\booksection{4. Byproducts}
\vskip-0.5truecm
\booksubsection{4.1 The electromagnetic coupling on the $Z$, $\bar{\alpha}_{\rm Q.E.D.}(M^2_{Z})$}

\noindent
With a simple change of integration kernel 
the previous analysis can be extended to evaluate the hadronic 
contribution to the QED running coupling, $\bar{\alpha}_{\rm Q.E.D.}(t)$, in 
particular on the $Z$ particle, $t=M^2_Z$; this is an important quantity that enters 
into precision evaluations of electroweak observables.
By using a dispersion relation one can write this  hadronic 
contribution at energy squared $t$ as 
$$
\Delta_{\rm had} \alpha(t) =-\dfrac{t\alpha}{3\pi}
\int_{4m^2_\pi}^\infty 
\dd s\;\dfrac{R(s)}{s(s-t)}
$$
where $R$ is as in (2.1b) and
the integral  has to be understood as a principal part integral.
Therefore, we can carry over all the work from the previous sections 
as in ref.~20, 
with the simple replacement   
$$K(s)\to-\dfrac{t\alpha}{3\pi}\dfrac{1}{s(s-t)}.$$
We find, to next to leading order in $\alpha$,
$$10^5\times\deltav_{\rm had}\alpha(M_Z^2)=
2\,742\pm12
\equn{(4.1)}$$
or, excluding the top quark contribution,
$$10^5\times\deltav_{\rm had}\alpha^{(5)}(M_Z^2)=
2\,749\pm12.
$$
Adding the known pure QED corrections,
the running QED coupling, in the momentum scheme, is
$$\bar{\alpha}_{\rm Q.E.D.}(M_Z^2)=
\frac{1}{128.962\pm0.016}.
\equn{(4.2)}$$
The difference with the result in ref.~20 is minute. 
(4.2) may be compared with a recent calculation of Hagiwara et al.,\ref{21} 
who, using only experimental $e^+e^-$ annihilation data at low energy 
find
$$\bar{\alpha}_{\rm Q.E.D.}(M_Z^2)=
\frac{1}{128.954\pm0.031}.
\equn{(4.3)}$$
While the central value is compatible with ours, the error 
in (4.3) is twice as large. This shows clearly 
the advantage of a thorough use of analyticity as well as of combining several sets of data 
(we will discuss this further in \sect~6, in connection with the 
muon anomaly).

\booksubsection{4.2. The masses and widths of the rho,  the quadratic radius and 
second coefficient of the pion, and the scattering length and effective range parameter for the P wave 
in $\pi\pi$ scattering}

\noindent
As other byproducts of our analysis we can give very precise numbers for the 
masses and widths of the neutral and charged rho resonances,  for
 the quadratic radius and 
second coefficient of the pion form factor, and for the  the scattering length, 
$a^1_1$, and effective range parameter, $b^1_1$ (defined as in ref.~9), for the P wave 
in $\pi\pi$ scattering. 
In what regards the first, we have the following results: 
for the charged rho, and in \mev,
$$\matrix{\vphantom{\Big|}
&{\rm TY},\; \tau,e\pi,e^+e^-&{\rm GJ}\; 
(\tau\;\hbox{decay  only})
\cr m_{\rho^\pm}&774.0\pm0.4&775.4\pm0.4\cr
\gammav_{\rho^\pm}&147.7\pm0.7&149.3\pm0.4\cr
\chi^2/{\rm d.o.f.}&1.06&\sim1.8.
}
\equn{(4.3)}$$ 
For the neutral rho,
$$\matrix{\vphantom{\Big|}
&{\rm TY},\; e^+e^-, e\pi&{\rm TY},\; \tau,e\pi,e^+e^-&{\rm GJ}\; 
(e^+e^-\;\hbox{ only})
\cr m_{\rho^0}&773.1\pm0.6&773.2\pm0.4&772.95\pm0.7\cr
\gammav_{\rho^0}&141.7\pm1.2&146.0\pm0.8&147.9\pm0.7\cr
\chi^2/{\rm d.o.f.}&0.9&1.06&\sim1.7\cr
}
\equn{(4.4)}$$
Here TY are our results here, 
 and GJ refers to the results by Ghozzi and Jegerlehner.\ref{22}
Clearly, the masses are stable and well determined, the widths less so. 

The quadratic radius and second coefficient of the pion are defined by
$$F^2_\pi(t)\simeqsub_{t\to0}1+\tfrac{1}{6}\langle r^2_\pi\rangle t+c_\pi t^2,
\equn{(4.5)}$$
and our results allow a precise determination of both quantities. 
The same is true for $a^1_1$, $b^1_1$. 
We have, for $\pi^+\pi^-$,
$$\matrix{\vphantom{\Big|}
&{\rm TY}\;\hbox{[Only $e^+e^-$, $e\pi$]}&{\rm TY}\;\hbox{[Including $\tau$ decay]}&
{\rm Colangelo}\cr
\langle r^2_\pi\rangle\quad({\rm fm}^2)&0.423\pm0.003&0.432\pm0.001&0.435\pm0.005\cr 
c_\pi\;(\;{\gev}^{-4})&3.78\pm0.05&3.84\pm0.02&-\cr
a_1^1\;(m_\pi\equiv1)&(38.9\pm2.3)\times10^{-3}&(37.8\pm0.8)\times10^{-3}&(37.9\pm0.5)
\times10^{-3}\cr
b_1^1\;(m_\pi\equiv1)&(4.1\pm0.7)\times10^{-3}&(4.74\pm0.09)\times10^{-3}&(5.67\pm0.13)
\times10^{-3}.
\cr}
\equn{(4.6)}$$
TY is from our results here and ``Colangelo"  from refs.~8,~9. 
The discrepancy between the result for $b^1_1$ from the pion form factor 
and that in ref.~9 had already been noted in ref.~10.
The numbers coming from our calculations for  $a^1_1$, $b^1_1$, 
{\sl including} tau decay data, agree much better than those not including it 
with results from pion-pion scattering, either from phase shifts analyses or 
using the Froissart--Gribov representation,\ref{10} 
which is another reason for preferring the results including tau decays.

The values of the parameters of our fits are
$$\eqalign{
e^+e^-,&\,\;e\pi:\cr
c_1=&\,0.26\pm0.04,\quad c_2=0.19\pm0.13,\cr
b_0=&\,1.106\pm0.009,\quad b_1=0.23\pm0.10
}
\equn{(4.7)}$$
and
$$\eqalign{
e^+e^-,&\,\;e\pi,\;\tau:\cr
c_1=&\,0.24\pm0.01,\quad c_2=-0.18\pm0.03,\cr
b_0=&\,1.074\pm0.006,\quad b_1=0.13\pm0.04.
}
\equn{(4.8)}$$
The numbers $b_i$ in (4.8) correspond to $\pi^+\pi^-$; 
they are the ones we have used to calculate $a_\mu(s\leq0.8\,\gev^2)$.
 For $\pi^0\pi^+$ we would have,
$$b_0=1.064\pm0.006,\quad b_1=0.13\pm0.03\quad[\pi^0\pi^+],
\equn{(4.9)}$$
and the $c_i$ are as in (4.8). 
The corresponding scattering length and effective range parameter are, in units of
$m_{\pi^+}$,
$$a_1^1=(37.8\pm0.8)\times10^{-3},\quad b_1^1=(4.78\pm0.09)\times10^{-3}\quad[\pi^0\pi^+].
\equn{(4.10)}$$

\booksection{5. Comparison of our theoretical calculations with experiment for $a_\mu$}

\noindent
We return to the magnetic moment of the muon, 
and present,  in Table~I,  a summary of our   
results  for $a{\rm (Hadr.)}$. 
In this Table we have added our old result from 2002, ref.~6, and the result of 
a recent evaluation by
Jegerlehner,\ref{21}  in which theory is kept to a minimum; for example, only data are used
for 
$s<0.8\,\gev^2$, and QCD is taken valid only for energies above 13 \gev. 
The main interest of this type of calculation lies in its role 
as control of the calculations where a more comprehensive use 
of theory (as well as extra experimental information) is made.
In the Table we also include the recent experimental value of ref.~2.  
For our evaluations here, 
we have separated explicitly the estimated errors due to 
radiative corrections, and the light-by-light scattering diagram.

\bigskip
\midinsert{
\setbox1=\vbox{\petit\offinterlineskip\hrule
\halign{
&\vrule#&\strut\hfil#\hfil&\quad\vrule\quad#&\strut\hfil#\hfil&
\quad\vrule#&\strut\hfil$\;$#\hfil\cr 
height1mm&\omit&&\omit&&\omit&\cr
&\vphantom{\Big|}  &&$10^{11}\times a_\mu(s\leq0.8\,\gev^2)$&&$10^{11}\times a_\mu(\hbox{Hadr.})$& \cr
\cr
\noalign{\hrule}
&\vphantom{\Big|} Jegerlehner &&\hfil --\hfil&
&$6\,840\pm94$ &\cr
\noalign{\hrule}
&\vphantom{\Big|}  TY (2002), $e^+e^-$, $e\pi$, $\tau$ &
&$4\,774\pm51$ &&$6\,993\pm69$\phantom{L}& \cr
\noalign{\hrule}
&\vphantom{\Big|}  TY, $e^+e^-$, $e\pi$  &&$4\,715\pm32\pm10\,({\rm rad.})$&
&$6\,935\pm50\pm10\,({\rm rad.})\pm30\,(\ell\times\ell)$ \phantom{L}&\cr
\noalign{\hrule}
&\vphantom{\Big|}  TY, $e^+e^-$, $e\pi$, $\tau$ &
&$4\,798\pm31\pm10\;({\rm rad.})$ &&$7\,018\pm49\pm10\,
({\rm rad.})\pm30\,(\ell\times\ell)$\phantom{L}& \cr
\noalign{\hrule}
& \vphantom{\Big|} Experiment && &&$7\,209\pm60$&\cr
 height1mm&\omit&&\omit&&\omit&\cr
\noalign{\hrule}}
\vskip.05cm
}
\centerline{\box1}
\medskip\centerline{\sc Table I}
\medskip
\setbox2=\vbox{\hsize=0.99\hsize \noindent{\petit Contributions to the rho region,
 and to the hadronic part of
the muon anomaly. 
 Jegerlehner: ref.~22. TY (2002): ref.~6. TY: this article.}
\medskip
\centerrule{4truecm}}
\centerline{\box2} 
}\endinsert

From  Table~I it is clear that there is reasonable agreement 
among the various theoretical
determinations, but there is a definite distance between the central values 
from theory and experiment, at a level between $2.3\,\sigma$ and  $3.3\,\sigma$,
 if we add quadratically ``rad" and ``$\ell\times\ell$" 
errors to the other ones.   
In the remaining of this section we will discuss 
possible reasons for this discrepancy.

An obvious reason would be new physics; 
we will not discuss this 
here, since it lies outside the scope of the present paper, and send the interested reader to 
the   hundreds of papers that have been written discussing this possibility.

A  second  reason is, of course, a displacement of the experimental result. 
Since the experimental number in \equn{(1.1)} comes basically from only 
{\sl one} experiment, it could happen that 
an independent determination would move  
it to better agreement with the  results of the theoretical evaluations. 

  And a third possibility is that the central
values of some of
 the theoretical evaluations presented here 
are displaced with respect to the true values. 
The more obvious place where such a displacement may occur is the evaluation of 
$a(\hbox{$\ell\times \ell$})$. 
The two approximations used to evaluate this contribution
 do not have overlapping ranges of validity; there 
is a wide region, when the virtualities of some (one) of 
the intermediate photons are small, and at the
 same time other (others) are large, 
where neither the one-pion or the constituent quark approximations 
need to be valid. 
In fact, the only certain result we have on this piece is the coefficient of the leading  
chiral logarithm ($\log^2 m_\pi$) given in ref.~23, which is of little 
practical use. However, it is not easy to see how one
could get the  large values necessary for theory and experiment to overlap.
In a recent calculation, using methods somewhat 
different to previous ones, Melnikov and Vainshtein\ref{24} find 
 $10^{11}\times a(\hbox{$\ell\times \ell$})=
136\pm25$. Although this is  larger by a bit more than one sigma  ($44\times\,10^{-11}$)
 than the result quoted in
(3.3),  it is not sufficient to  remove the discrepancy.

Finally, we may have a coincidence of several of the effects
 mentioned here, with the bad luck that 
they add.

\booksection{6. Comparison with other recent calculations and\hb concluding remarks}

\noindent
Our analysis shows that, to get a precise value of $a_\mu$,
 it is certainly necessary to profit from the
existence of methods  that allow us to make full use of theory, in particular for 
fitting the pion form factor:  the use of robust theory 
in the fits produces   robust results. 
It is not very consistent to use analyticity and unitarity to write the 
representation (2.1), and refuse to use exactly the same ingredients to 
improve the knowledge of $F_\pi$. 
In this sense, it is also important to take into account the data on $F_\pi(s)$ 
for spacelike $s$, i.e., from $e\pi$ scattering. 
This has been deemed inappropriate by some authors because they 
(may) contain systematic errors. 
However, if one refrained from taking into account data afflicted by systematic 
errors, there would be no data one could use.
Systematic errors can and should be taken into account as shown, in this 
particular instance, in ref.~6 and in the present article. 

In the same vein, we believe that tau decay data can and should be used, 
in spite of the fact that the data of the various tau decay experiments differ in 
some energy regions by more than one standard deviation --doubtlessly because of systematic 
errors, as is obvious from  (2.11c). 
This is particularly important because the errors given in 
Eq.~(2.8a) for $a_\mu^{(2)}(s\leq0.8\,\gev^2)$ are deceptively 
small. 
As we already commented, the error per experimental point for $e^+e^-\to\pi\pi$,  
fitting only $e^+e^-$, $e\pi$ data, is  
$89/113\simeq0.79$, clearly smaller than unity; while, even imposing tau decay information, 
the error per experimental point is (as reported in Eq.~(2.11c)) 
only of $108/113\simeq0.97$, perfectly acceptable. 
This means that  acceptable fits --like, indeed, the one obtained by us 
using also tau decay data-- 
can be found outside the nominal error bars in Eq.~(2.8a).\fnote{One should, 
however, not forget that the value we needed for the 
tau decay normalization error, $(-1.4\pm0.5)\%$ (cf.~Eq.~(2.11b)), or  
 $(-1.2\pm0.5)\%$ if including the estimate (2.16) for $\delta_\gamma$, is slightly 
larger than the expected normalization error, as given by the Particle Data Tables,
$0.7\%$.}

Likewise, when data on $F_\pi$ from the processes  $e^+e^-\to\pi^+\pi^-+\gamma$
are forthcoming, they should be incorporated into the analysis: 
the safest way to get rid of systematic errors is to combine data of 
various, independent experiments, so that the various independent systematic errors average out. 
The {\sl gain}, both in accuracy and robustness that follows from our methods can 
perhaps  be seen more clearly if we compare them with other recent evaluations,\ref{21,25,26} 
something that we do in Table~II (where we do not include the results of 
ref.~22, already discussed before). 
Here the stability of our 2002 results against including new, more precise $e^+e^-$ data, 
contrasts with the variations in the other determinations, is spite of the 
fact that our error is substantially smaller.

\bigskip
\midinsert{
\setbox1=\vbox{\petit\offinterlineskip\hrule
\halign{
&\vrule#&\strut\hfil#\hfil&\quad\vrule\quad#&\strut\hfil#\hfil&
\quad\vrule#&\strut\hfil$\;$#\hfil\cr 
height1mm&\omit&&\omit&&\omit&\cr
&\vphantom{\Big|}  &&$10^{11}\times a_\mu^{\rm had,\;LO+\gamma}$&&$10^{11}\times a_\mu$& \cr
\cr
\noalign{\hrule}
&\vphantom{\Big|} Ezhela\quad[$e^+e^-$]  &&\hfil $6\,996\pm89$\hfil&
& $116\,591\,835\pm96$\phantom{L}&\cr
\noalign{\hrule}
&\vphantom{\Big|} Hagiwara\quad[$e^+e^-$]  &&\hfil $6\,924\pm64$\hfil&
&$116\,591\,763\pm74$\phantom{L} &\cr
\noalign{\hrule}
&\vphantom{\Big|} Davier (a)\quad[$e^+e^-$]  &&\hfil $6\,847\pm70$\hfil&
& $116\,591\,693\pm78$\phantom{L}&\cr
\noalign{\hrule}
&\vphantom{\Big|} Davier (b)\quad[$e^+e^-$]  &&\hfil $6\,963\pm72$\hfil&
&$116\,591\,809\pm80$\phantom{L} &\cr
\noalign{\hrule}
&\vphantom{\Big|} Davier (b)\quad[$\tau$]  &&\hfil $7\,110\pm58$\hfil&
&$116\,591\,956\pm68$\phantom{L} &\cr
\noalign{\hrule}
&\vphantom{\Big|}  TY (2002), $e^+e^-$, $e\pi$, $\tau$ &
&$7\,002\pm66$ &&$116\,591\,849\pm69$\phantom{L}& \cr
\noalign{\hrule}
&\vphantom{\Big|}  TY, $e^+e^-$, $e\pi$, $\tau$ &
&$7\,027\pm49$ &&$116\,591\,889\pm58$\phantom{L}& \cr
\noalign{\hrule}
& \vphantom{\Big|} Experiment && &&$116\,592\,080\pm60$\phantom{L}&\cr
 height1mm&\omit&&\omit&&\omit&\cr
\noalign{\hrule}}
\vskip.05cm
}
\centerline{\box1}
\medskip\centerline{\sc Table II}
\medskip
\setbox2=\vbox{\hsize=0.99\hsize \noindent{\petit The lowest order hadronic part of
the muon anomaly, including photon corrections [Eq.~(3.9)], and the full $a_\mu$. 
 Hagiwara: ref.~21. Ezhela: ref~25. Davier, (a) and (b): ref.~26,
 (a) and (b).  TY (2002): ref.~6. TY: this article.}
\medskip
\centerrule{4truecm}}
\centerline{\box2} 
}\endinsert

In short: fitting to the theoretical expressions instead of integrating
directly the data, allows us  to compare
the different data samples among them in a fully quantitative manner. 
We believe that only using this kind of quantitative comparisons one can
decide if the suggested discrepancies are meaningful or only apparent, and,
if meaningful, if they are due to systematic errors  or to
physics. The fact is that one can fit all the $e\pi$, $e^+e^-\to\pi\pi$ and tau data with
a \chidof\  essentially 1. This result is a non-trivial improvement on previous work.

It is, however, not clear to us that one can improve the results using 
$\pi\pi$ scattering data. 
If we include in the fit  
the {\sl experimental} numbers for $\delta^1_1$, 
the value of $a_\mu$ increases  by $8\times10^{-11}$; 
but this is not necessarily more precise than the result without including this 
information.
 The systematic errors of  $\delta^1_1$, 
due to the fact that one does not scatter real pions (and 
thus one has to rely on models), are larger than the errors in our calculation. 
And if, like Colangelo and collaborators,\ref{8,9} 
we input  $\delta^1_1$ from theoretical analyses (Roy 
equations and chiral perturbation theory), 
one is depending on determinations whose accuracy has been challenged\ref{10} and 
is, very likely, too optimistically estimated. 

Apart from this, to get real improvement in the theoretical predictions
 for the quantity $a_\mu({\rm Hadr.})$ it would be,  first of  all,
 necessary to remove the sources of
uncertainty mentioned at the end of \sect~5.  Of these,  the one stemming from 
$\ell\times \ell$ is unlikely to be removed in a satisfactory manner; 
and we have also a problem (although less important numerically) 
with electroweak radiative corrections to tau decay. 
If we treat them by considering the pions as elementary, and factoring out $F_\pi$, 
then Sirlin's theorem\ref{12} implies that, for $\tau\to\nu\pi\pi$, the 
logarithmic piece, $\log M_Z/m_\tau$ cancels out. 
It, however, does not cancel if we consider that, at short distances, 
the decay is really $\tau^-\to\nu \bar{u}d$. 
This means that the model with elementary pions fails, for this case 
of tau decay, and thus that the chances of removing the uncertainties due to 
lack of accurate knowledge of the radiative corrections here are remote.

As a final comment, we would like again to bring attention to the 
mismatch between the experimental and theoretical values 
for $a_\mu$; although not yet definite proof of failure of the 
standard model, it cannot be  ignored.

\vfill\eject
\booksection{Acknowledgements}

\noindent
We are  grateful to Fred Jegerlehner for discussions --oral and e-mail-- 
that have helped a lot to clarify important issues.

\booksection{References}

\item{1 }{The  QED calculations of $a_e$, 
and (some of) those of $a_\mu$,
 with references, may be found in the review of 
V.~W.~Hughes and T.~Kinoshita, {\sl Rev. Mod.
Phys.} {\bf 71}, S133 (1999).}
\item{2 }{H.N. Brown et al., {\sl Phys. Rev. Letters} {\bf 86}, 2227 (2001); 
G.~W.~Bennett et al.,  {\sl Phys. Rev. Letters} {\bf 89}, 101804 and (E)~129903 (2002); 
G.~W.~Bennett et al.,
hep-ex/0401008 v3.}
\item{3 }{For the electromagnetic corrections, 
see the text  by 
 T.~Kinoshita et al., {\sl Quantum Electrodynamics}, World
Scientific, 1990 and T.~Kinoshita and M.~Nio, hep-ph/0402206, whose numbers we take. 
Weak corrections were 
first calculated by 
R.~Jackiw and S.~Weinberg, {\sl Phys. Rev.} {\bf D5}, 2396 (1972); 
I.~Bars and M.~Yoshimura, {\sl ibid.} {\bf D6}, 374 (1972); 
K.~Fujikawa, B.~W.~Lee and A.~I.~Sanda, {\sl ibid} 
2932 (1972). The hadronic weak  corrections have been recently reconsidered  M.~Knecht et al., {\sl
JHEP} {\bf 0211}, 003 (2002), and a detailed 
 reassessment of all electroweak contributions is that by 
A.~Czarnecki, W.~J.~Marciano and A.~Vainshtein,
 {\sl Phys. Rev.}  {\bf D67}, 073006 (2003).}
\item{4 }{\/Novosibirsk, $\rho$ region: L.~M.~Barkov et al., {\sl Nucl. Phys.} {\bf B256}, 365 (1985); 
R.~R.~Akhmetshin et al., Budker INP 99-10 (1999) [hep-ex/9904027], superseded by  R.~R.~Akhmetshin et al.,
{\sl Phys. Letters} {\bf B527}, 161
(2002). The details of the calculations of the radiative corrections to 
$e^+e^-\to\pi\pi$ scattering, necessary to extract the 
pion form factor from data, are given in P.~Singer, {\sl Phys. Rev.} {\bf130}, 2441 (1963) and Erratum, 
{\bf 161}, 1694 (1967) and A.~B.~Arbuzov et al., {\sl J. High Energy Phys.} 
{\bf 10}, 1 and 6 (1997).}
\item{5 }{Novosibirsk, $\omega$ and $\phi$ region, $KK$ and $3\pi$: 
R. R. Akhmetshin et al., {\sl Phys. Letters} {\bf B466}, 385 and 392 (1999); 
ibid., {\bf B434},  426 (1998) and ibid., {\bf B476},  33 (2000); 
 hep-ex/0308008 (2003). 
M.~N.~Achasov et al., {\sl Phys. Rev.} {\bf D63}, 072002 (2001);    
M.~N.~Achasov et al., {\sl Phys. Letters} 365 {\bf B462} (1999) and   
 Preprint Budker INP 98-65 (1998) [hep-ex/9809013]. 
$\phi\to2\pi$:  M.~N.~Achasov et al., {\sl Phys. Letters} {\bf B474}, 188 (2000).}
\item{6 }{J.~F.~de~Troc\'oniz and F.~J.~Yndur\'ain, {\sl Phys. Rev.}  {\bf D65}, 093001 (2002).}
\item{7 }{S. R. Amendolia et al., {\sl Nucl. Phys.} {\bf B277}, 168 (1986).}
\item{8 }{H. Leutwyler, hep-ph/0212324; 
G.~Colangelo, hep-ph/0312017.}
\item{9 }{G.~Colangelo, J. Gasser, and H.~Leutwyler
 {\sl Nucl. Phys.} {\bf B603},  125, (2001).}
\item{10 }{J. R. Pel\'aez and  F.~J.~Yndur\'ain,  { Phys. Rev.} {\bf D68}, 074005 (2003) 
and FTUAM  03-19 (2003)
[hep-ph/0312187].} 
\item{11 }{ALEPH: R. Barate et al., {\sl Z. Phys.} {\bf C76}, 15 (1997); OPAL:
K.~Ackerstaff et al., {\sl Eur. Phys. J.} {\bf C7}, 571 (1999); CLEO: S.~Anderson et al., {\sl Phys. Rev.},
 {\bf D61}, 112002 ((2000).}
\item{12 }{A. Sirlin, {\sl Nucl. Phys.} {\bf B196}, 83 (1982); W.~J.~Marciano and 
A.~Sirlin, {\sl Phys. Rev. Lett.} {\bf 61}, 1815 (1988).}
\item{13 }{$KK$\/: {P. M. Ivanov et al.}, {\sl Phys. Letters} {\bf 107B}, {297} (1981); $3\pi$: 
A.~Cordier et al., {\sl Nucl. Phys.} {\bf B172}, 13 (1980); $4\pi$ and more: 
G.~Cosme et al., {\sl Nucl. Phys.} {\bf B152}, 215 (1979). [For a review, see  
S. Dolinsky et al., {\sl Phys. Reports} {\bf C202}, 99 (1991), and other work quoted there]. 
All these references 
give results for energies below $s=2\,\gev^2$. 
Between $2$ and $9\;\gev^2$, see 
C. Bacci et al., {\sl Phys. Letters}  {\bf 86B}, 234 (1979). 
At the $\bar{c}c$ threshold region: J. Z. Bai et al., {\sl Phys. Rev. Lett.}, {\bf 88}, 1010802 
(2002).}
\item{14 }{For the QCD expressions: K.~G.~Chetyrkin, A.~L.~Kataev and 
F.~V.~Tkachov, 
{\sl Phys. Letters} {\bf B85}, 277 (1979); M.~Dine and J.~Sapiristein, 
{\sl Phys. Rev. Lett.} {\bf 43}, 668 (1979); W.~Celmaster and R.~Gonsalves, {\sl Phys. Rev. Lett.} {\bf 44},
560 (1980);  S.~G.~Gorishny, A.~L.~Kataev and S.~A.~Larin, {\sl Phys. Letters} {\bf B259}, 144 (1991); 
L.~R.~Sugurladze, and M.~A.~Samuel, {\sl Phys. Rev. Lett.} {\bf 66}, {560} (1991).  
$s$ quark mass: S.~Chen et al., {\sl Eur. J. Phys.} {\bf C22}, {31} (2001); $b,\,c$ quark masses: 
 A. Pineda and
F.~J.~Yndur\'ain, {\sl Phys. Rev.} {\bf D58}, 094022 (1998)  and {\sl Phys. Rev.}  {\bf D61}, 077505 (2000).
$\alpha_s$:  F.~Le~Diberder and A.~Pich, {\sl Phys. letters} {\bf B289}, 165 (1992);
J.~Santiago and F. J. Yndur\'ain,  {\sl Nucl. Phys.} {\bf B563}, 45 (1999) 
and  {\sl Nucl. Phys.} {\bf B611}, 447 (2001). 
For a review, see S. Bethke, {\sl J. Phys.} {\bf G26}, {R27} (2000) and, for the 
direct measurement of $\alpha_s$ on the Z,
 D. Strom, ``Electroweak measurements on the $Z$ resonance", 
Talk presented at the 5th Int. Symposium on Radiative Corrections, RadCor2000, 
Carmel, Ca., September 2000.}
\item{15 }{S. Narison, {\sl Phys. Letters} {\bf B513}, 53 (2001) and (E) {\bf B526}, 414
(2002).}
\item{16 }{J. Bijnens, E. Pallante and J. Prades, {\sl Nucl. Phys.} {\bf B474},
 379 (1996); 
 M.~Hayakawa, T.~Kinoshita and A.~I.~Sanda, {\sl Phys. Rev.} {\bf D54},
 3137 (1996); 
M.~Hayakawa and T.~Kinoshita, {\sl Phys. Rev.} {\bf D57}, 465 (1998).}
\item{17 }{M. Knecht and A. Nyffeler, {\sl Phys. Rev.} {\bf D65}, 073034 (2002).}
\item{18 }{B. Krause, {\sl Phys. Letters} {\bf B390}, {392} (1997)}
\item{19 }{{P. Singer, }{\sl Phys. Rev.}{\bf130}, {2441} ({1963}) and (E),  
{\bf 161}, 1694 (1967).}
\item{20 }{J.~F.~de~Troc\'oniz and F.~J.~Yndur\'ain, {\sl Phys. Rev.}  {\bf D65}, 093002
(2002).}
\item{21 }{K. Hagiwara et al., {\sl Phys. Rev.} {\bf D69}, 093004 (2004).}
\item{22 }{F.~Jegerlehner, Proc. Int. Frascati Conf., 2003  [hep-ph/0310234]; 
S.~Ghozzi and F.~Jegerlehner, {\sl Phys. Letters} {\bf B583}, 222 (2004).}
\item{23 }{Knecht et al., {\sl Phys. Rev. Lett.} {\bf 88}, 071802 (2002).}
\item{24 }{K.~Melnikov and A. Vainshtein, FTPI-MINN-03-36 (hep-ph/0312226).}
\item{25 }{V. V. Ezhela, S. B. Lugovsky and O. V. Zenin, IHEP-2003-35 
(hep-ph/0312114).}
\item{26 }{(a) M. Davier et al., Eur. Phys. J. {\bf C27}, 497 (2003); (b) 
 M. Davier et al., Eur. Phys. J. {\bf C31}, 503 (2003).}

\bye